\documentclass[paper]{JHEP3} 

\JHEPspecialurl{http://jhep.sissa.it/JOURNAL/JHEP3.tar.gz}

\usepackage{epsfig,multicol,bbm}

\newcommand\fverb{\setbox\fverbbox=\hbox\bgroup\verb}
\newcommand\fverbdo{\egroup\medskip\noindent%
			\fbox{\unhbox\fverbbox}\ }
\newcommand\fverbit{\egroup\item[\fbox{\unhbox\fverbbox}]}
\newbox\fverbbox

\newcommand{\be}{\begin{equation}}
\newcommand{\ee}{\end{equation}}
\newcommand{\bea}{\begin{eqnarray}}
\newcommand{\eea}{\end{eqnarray}}

\def\a{\alpha}

\def\g{\gamma}
\def\G{\Gamma}

\def\la{\lambda}

\def\s{\sigma}

\def\f{\phi}

\def\ep{\epsilon}
\def\vep{\varepsilon}
\def\th{\theta}

\newcommand{\eg}{{\it e.g.,}\ }
\newcommand{\ie}{{\it i.e.,}\ }

\title{Supersymmetric codimension-two branes and $U(1)_R$ mediation in 6D gauged supergravity}

\author{Hyun Min Lee\\
	Department of Physics, Carnegie Mellon University,
Pittsburgh, PA 15213, USA.\\
	E-mail: \email{hmlee@andrew.cmu.edu}}

\preprint{March 2008}	

\abstract{We construct a consistent supersymmetric action for brane chiral and vector multiplets in a six-dimensional chiral gauged supergravity. A nonzero brane tension can be
accommodated by allowing for a brane-localized Fayet-Iliopoulos term proportional to the brane tension.
When the brane chiral multiplet is charged under the bulk $U(1)_R$, 
we obtain a nontrivial coupling to the extra 
component of the $U(1)_R$ gauge field strength and a singular scalar self-interaction term. Dimensionally reducing to 4D on a football supersymmetric solution, we discuss the implication of such interactions for obtaining the $U(1)_R$ D-term in the 4D effective supergravity. By assuming the bulk gaugino condensates and nonzero brane F- and/or D-term for the uplifting potential,
we have all the moduli stabilized with a vanishing cosmological constant. 
The brane scalar with nonzero $R$ charge then gets a soft mass of order the gravitino mass. 
The overall sign of the soft mass squared depends on the sign of the $R$ charge as
well as whether the brane F- or D-term dominates.  
}


\begin{document} 


\section{Introduction}

Supersymmetry(SUSY)\cite{reviewonsusy} has been around us as one of the most promising candidates 
for physics beyond the Standard Model. When the SUSY breaking occurs at the electroweak scale,
it can be a solution to the gauge hierarchy problem, thanks to the cancellation of the
quadratic divergences to the Higgs mass between the
SM particles and their superpartners. If the SUSY breaking is parametrized in terms of
the soft mass parameters that respect the SM gauge symmetry, over 100
additional parameters would lead to unacceptably large FCNCs and CP violations. 
In gravity mediation\cite{gravmed} where the hidden sector SUSY breaking is transmitted to the visible sector by gravity only, the weak scale soft mass parameters can be naturally generated.
However, the almost flavor-universal soft masses are unexplained in this context because one cannot forbid the flavor-dependent contact interactions between the visible and hidden sectors by any known
symmetry. 
Therefore, several alternative mechanisms of SUSY mediation giving the flavor-universal 
soft masses have been suggested and discovered:
gauge mediation\cite{gaugemed}, anomaly mediation\cite{anomalymed}, gaugino mediation\cite{gauginomed}, etc.

For recent years, there have been a plenty of interest in the flux compactifications for the SUSY phenomenology, in particular, in the context of the KKLT-type compactifications\cite{kklt}, due to the fact that fluxes combined with non-perturbative effect can fix all the moduli of the extra dimensions endowed from string theory and the soft mass parameters have a distinct pattern compared to the ones obtained in the other SUSY mediations, that leads to the so called anomaly-modulus mixed mediation or the mirage mediation\cite{mirage}. In the present era that the Large Hadron Collider(LHC) is turning on soon,
it is compelling and very important to identify the distinguishable features of the reasonable  and accessible SUSY mediation mechanisms.

The model that we are considering is the Salam-Sezgin supergravity\cite{SS} 
where the $U(1)_R$ subgroup of the bulk $R$ symmetry is gauged by a bulk vector multiplet.
Due to the $U(1)_R$ gauging, there appears a nonzero positive scalar potential for the dilaton, in contrast to the 5D gauged supergravity.
In this model, due to the cancellation between the dilaton potential and the $U(1)_R$ gauge flux, the 4D Minkowski space with factorizable extra dimensions compactified on a sphere was obtained. The solution has been generalized rather recently to the unwarped or warped 4D Minkowski solutions with nonzero brane tensions\cite{football,branesol,gibbons,leelud}, with the hope of achieving a self-tuning of the cosmological constant on a codimension-two brane\cite{selftune}. 
In this case, nonzero brane tensions need to be introduced
at the conical singularities that are caused by the deficit angles. 
The brane tensions are, however, regarded as breaking the bulk SUSY explicitly at the action level.
Therefore, most recently, the SUSY action for the case with a brane tension has been constructed
by adding localized FI terms and localized corrections to the Chern-Simons term in the field strength
for the Kalb-Ramond field as well as by modifying the fermionic SUSY transformations\cite{susysolution}.
Consequently, it has been shown that the unwarped football solution with arbitrary brane tensions is a new SUSY background solution preserving the 4D ${\cal N}=1$ SUSY\cite{susysolution}. 

In this paper,
by extending the previous result for the SUSY action for a brane tension action of Ref.~\cite{susysolution},  
we first construct a consistent SUSY action in the presence of brane multiplets in the Salam-Sezgin supergravity by following the Noether method undertaken in 6D ungauged supergravity\cite{fll}. In this process, we need to add the localized terms depending on the brane multiplets to the
field strengths and modify the fermionic SUSY transformations and the SUSY/gauge
transformations of the Kalb-Ramond field.
Consequently, as in the ungauged supergravity\cite{fll}, we find that the kinetic term for the brane chiral multiplet has a nontrivial dilaton coupling while the gauge kinetic function for the brane vector multiplet is trivial. Moreover, the brane-induced F- and D-terms have nontrivial moduli couplings.
When the brane chiral multiplet is charged under the $U(1)_R$, we also obtain
a nontrivial coupling of the brane scalar to the gauge field strength as well as a singular
scalar self interaction.


For the SUSY background solution with football-shaped extra dimensions, we consider the low energy action with light bulk and brane modes in the 4D
effective supergravity.
The $U(1)_R$ gauge symmetry appears anomalous as the bulk Green-Schwarz counterterm generates the 4D $U(1)_R$ anomalies for a nonzero gauge flux and a localized FI term.  
Moreover, the bulk $U(1)_R$ vector multiplet gets a mass of order the 4D Planck scale by a Green-Schwarz mechanism. From the effective $U(1)_R$ D-term potential, we have fixed only one of two moduli, i.e. the $T$ modulus, due to the interplay between the field-dependent and constant $U(1)_R$ FI terms. By assuming that there appears an additional scalar potential due to two bulk gaugino condensates generated below the compactification scale in an extend bulk theory and consequently adding the brane F- and/or D-term as the uplifting potential, we show that the
remaining $S$ modulus is also stabilized at the Minkowski vacuum. 
Due to the shift of the VEV of the $T$ modulus from the one obtained only by the $U(1)_R$ D-term, 
we show that a brane scalar field with nonzero $R$ charge can get a nonzero soft mass in comparable size to the gravitino mass. We dub this new possibility ``$U(1)_R$ mediation''.
The brane scalar soft mass does not depend on the $U(1)_R$ gauge coupling because the mass of the
$U(1)_R$ vector multiplet is also proportional to the $U(1)_R$ gauge coupling.
The overall sign of the soft mass depends on whether the brane F-term or D-term dominates.

The paper is organized as follows.
We start with describing the bulk Salam-Sezgin supergravity and then present the SUSY action for the chiral and vector multiplets living on the
conical branes and the brane-localized D- and F-terms with nontrivial moduli dependence. 
Next we review on the recently found SUSY brane solution and continue to
derive the low energy effective action for light bulk and brane fields in the same SUSY brane background to identify the corresponding 4D effective supergravity.
We also discuss on the moduli stabilization and the SUSY breaking in the presence of the bulk gaugino condensates and nonzero brane F- and/or D-term uplifting potentials.
Finally, the conclusion is drawn.

\section{Model setup}

The six-dimensional Salam-Sezgin supergravity \cite{SS} consists
of gravity coupled to a dilaton field $\f$, a Kalb-Ramond(KR) field $B_{MN}$, along with the 
SUSY fermionic partners, the gravitino $\psi_M$, the dilatino $\chi$.
Moreover, it also contains a bulk $U(1)_R$ vector multiplet $(A_M,\lambda)$ that gauges the $R$-symmetry of six-dimensional supergravity. All the bulk fermions are 6D Weyl.

The complete bulk Langrangian\cite{SS} is given (up to four fermion terms) by
 \bea e^{-1}_6{\cal L}_{\rm
bulk}&=&R-\frac{1}{4}(\partial_M\phi)^2-\frac{1}{12}
e^{\phi}G_{MNP}G^{MNP}
-\frac{1}{4}e^{\frac{1}{2}\phi}F_{MN}F^{MN}-8g^2 e^{-\frac{1}{2}\phi} \nonumber \\
&&+{\bar\psi}_M\Gamma^{MNP}{\cal D}_N\psi_P+
{\bar\chi}\Gamma^M{\cal D}_M\chi
+{\bar\lambda}\Gamma^M{\cal D}_M\lambda\nonumber\\
&&+\frac{1}{4}(\partial_M\phi)({\bar\psi}_N\Gamma^M\Gamma^N\chi +
\bar{\chi}\Gamma^N\Gamma^M\psi_N)\nonumber \\
&&+\frac{1}{24} e^{\frac{1}{2}\phi}G_{MNP}(\bar{\psi}^R\Gamma_{[R}\Gamma^{MNP}\Gamma_{S]}\psi^S
+\bar{\psi}_R\Gamma^{MNP}\Gamma^R\chi \nonumber \\
&&\quad-\bar{\chi}\Gamma^R\Gamma^{MNP}\psi_R
-{\bar\chi}\Gamma^{MNP}\chi+\bar{\lambda}\Gamma^{MNP}\lambda) \nonumber \\
&&-\frac{1}{4\sqrt{2}}e^{\frac{1}{4}\phi}F_{MN}({\bar\psi}_Q\Gamma^{MN}\Gamma^Q
\lambda+\bar{\lambda}\Gamma^Q\Gamma^{MN}\psi_Q+{\bar\chi}\Gamma^{MN}\lambda-\bar{\lambda}\Gamma^{MN}\chi) \nonumber \\
&&+i\sqrt{2}g e^{-\frac{1}{4}\phi}({\bar\psi}_M\Gamma^M\lambda
+\bar{\lambda}\Gamma^M\psi_M-{\bar\chi}\lambda+\bar{\lambda}\chi). \label{bulkaction}
\eea The field strengths of the gauge and Kalb-Ramond(KR) fields
are defined as \bea
F_{MN}&=& \partial_M A_N - \partial_N A_N, \label{fmn} \\
G_{MNP}&=& 3\partial_{[M}B_{NP]} +\frac{3}{2} F_{[MN}A_{P]},
\label{gmn} \eea and satisfy  the Bianchi identities \bea
\partial_{[Q}F_{MN]}&=&0, \label{obif}\\
\partial_{[Q}G_{MNP]}&=&\frac{3}{4}F_{[MN}F_{QP]}.\label{obig}
\eea For $\delta_\Lambda A_M =\partial_M \Lambda$ under the $U(1)_R$, the field strength for the KR field is made gauge invariant by allowing for $B_{MN}$ to transform as 
\be \delta_\Lambda B_{MN} =
-\frac{1}{2}\Lambda F_{MN}. \ee 
All the spinors have the same $R$ charge $+1$, so the covariant derivative of
the gravitino, for instance, is given by \be {\cal D}_M\psi_N =
(\partial_M+\frac{1}{4}\omega_{MAB}\Gamma^{AB}-igA_M)\psi_N. \ee

The local ${\cal N}=2$ SUSY transformations are (up to trilinear
fermion terms): \bea
\delta e^A_M &=& -\frac{1}{4}\bar{\vep}\Gamma^A\psi_M+{\rm h.c.}, \label{susyt1}\\
\delta\phi &=& \frac{1}{2}\bar{\vep}\chi +{\rm h.c.}, \label{susyt2}\\
\delta B_{MN} &=& A_{[M}\delta A_{N]}+\frac{1}{4}e^{-\frac{1}{2}\phi}
(\bar{\vep}\Gamma_M\psi_N-\bar{\vep}\Gamma_N\psi_M+\bar{\vep}\Gamma_{MN}\chi+{\rm h.c.}), \label{susyt3}\\
\delta\chi &=& -\frac{1}{4}(\partial_M\phi)\Gamma^M\vep
+\frac{1}{24}e^{\frac{1}{2}\phi}G_{MNP}\Gamma^{MNP}\vep, \label{susyt4} \\
\delta\psi_M &=& {\cal D}_M\vep +\frac{1}{48}e^{\frac{1}{2}\phi}G_{PQR}\Gamma^{PQR}\Gamma_M\vep, \label{susyt5} \\
\delta A_M &=& \frac{1}{2\sqrt{2}} e^{-\frac{1}{4}\phi}(\bar{\vep}\Gamma_M\lambda
+{\rm h.c.}), \label{susyt6} \\
\delta\lambda &=& \frac{1}{4\sqrt{2}} e^{\frac{1}{4}\phi}F_{MN}\Gamma^{MN}\vep
- i\sqrt{2}g\, e^{-\frac{1}{4}\phi}\vep. \label{susyt7}
\eea
The above spinors are chiral with handednesses
\be \G^7 \psi_M = +
\psi_M, \ \ \  \G^7 \chi = - \chi, \ \ \  \G^7 \la = + \la, \ \ \
\G^7 \varepsilon= + \varepsilon.
\ee
Taking into account that $\G^7=\s^3
\otimes {\bf 1}$ (see Appendix A), the 6D (8-component) spinors
can be decomposed to 6D Weyl (4-component) spinors as
\be \psi_M=
(\tilde{\psi}_M,0)^T, \ \ \  \chi =(0,\tilde{\chi})^T, \ \ \
\lambda= (\tilde{\la},0)^T, \ \ \ \varepsilon = (\tilde{\varepsilon},0)^T.
\ee
For later use, we decompose the 6D Weyl spinor
$\tilde{\psi}$ to $\tilde{\psi}=(\tilde{\psi}_{L},\tilde{\psi}_{ R})^T$,
satisfying $\gamma^5(\tilde{\psi}_{ L},0)^T=+(\tilde{\psi}_{ L},0)^T$ and
$\gamma^5 (0,\tilde{\psi}_{ R})^T=-(0,\tilde{\psi}_{ R})^T$. Henceforth we drop
the tildes for simplicity.

We can show that the action for the Lagrangian (\ref{bulkaction}) 
is invariant under the above SUSY transformations up to the
trilinear fermion terms and the Bianchi identities as follows,
\bea
\delta{\cal L}_{\rm bulk}&=&e_6\bigg[-\frac{1}{24}e^{\frac{1}{2}\phi}
\Big(\partial_S G_{MNP}-\frac{3}{4}F_{MN}F_{SP}\Big)\Big({\bar\psi}_R\Gamma^{RMNPS}\vep-{\bar\chi}\Gamma^{SMNP}\vep +{\rm h.c.}\Big) \nonumber \\
&&\quad +\frac{1}{4\sqrt{2}}e^{\frac{1}{4}\phi}\Big(\partial_Q F_{MN}{\bar\lambda}\Gamma^{QMN}\vep+{\rm h.c.}\Big)\bigg].\label{bulkvar}
\eea
Thus, as will be seen later, the SUSY variation of the brane action can be cancelled with the
bulk variation (\ref{bulkvar}) by modifying the Bianchi identities (\ref{obif}) and (\ref{obig}).

\section{Supersymmetric codimension-two brane actions}

We introduce a chiral multiplet with nonzero $R$-charge and a vector multiplet on the brane.
Then, by adopting the Noether method, we construct a consistent SUSY action for the brane multiplets 
that is invariant under the modified bulk SUSY variations.  
As a result, we show that the brane multiplets have nontrivial couplings to the bulk fields through the modified field strengths.
We also consider a supersymmetric brane-localized gravitino mass term.

\subsection{The $Z_2$ orbifold parities}

In order to project out half the bulk supersymmetries on the brane
and define an ${\cal N}=1$ brane SUSY, 
we assume an orbifold $Z_2$ symmetry around each brane.
If the local complex coordinate around the brane is $z$ (in
locally polar coordinates $z=re^{i \th}$), then the $Z_2$ symmetry
corresponds to \be z \leftrightarrow -z \ \ \ \ \ \ ({\rm or} \ \
\th \leftrightarrow \th + \pi). \ee 
In the case with two branes system, the warped vacua of
 \cite{gibbons} have an axially symmetric internal space. The above
 $Z_2$ symmetry about both branes present, is just a discrete subgroup
  of the axial symmetry. On the other
hand, for the general warped solutions with multiple branes
\cite{leelud}, we require the holomorphic function $V(z)$ in the
metric to satisfy the condition $|V(-z+z_i)|=|V(z-z_i)|$, where
$z_i$ is the $i$-th brane position.

We should then assign $Z_2$ parities to all bulk
fields as well as the SUSY variation parameters
$\vep_L$ and $\vep_R$. 
Being consistent with the bulk action and the SUSY transformation, 
we make a choice of parities for the fields and the SUSY variation parameter as \bea
{\rm even}&:&~ \psi_{\alpha L}, \ \psi_{a R}, \
\lambda_L, \
\chi_R, \ \varepsilon_L,\ A_\alpha,\ B_{\alpha\beta}, \ B_{ab}, \ \phi,\\
{\rm odd}&:&~  \psi_{\alpha R},\ \psi_{a L},\
\lambda_R,\
\chi_L, \ \varepsilon_R,\ A_a,\ B_{\alpha a}. \eea where
the gauge field, the Kalb-Ramond field and the gravitino have been
written with locally flat indices, \eg $A_A=e_A^{~M}A_M$, so that
the parity assignments do not depend on the coordinate system. It
is obvious that the above choice of parities forces
$\vep_R$ to vanish on the brane position.
Henceforth we denote the 4D Weyl spinor of each bulk fermion surviving on the brane by $\lambda_+$ and $\vep_+$, etc, satisfying $\gamma^5\lambda_+=\lambda_+$ and $\gamma^5\vep_+=\vep_+$, etc.

\subsection{The supersymmetric action for brane multiplets}

We consider a nonzero brane tension as well as brane matter multiplets: a brane chiral multiplet\footnote{The 4D chirality of the fermion in the brane chiral multiplet is taken to be right-handed in contrast to the $Z_2$-even gravitino and the $Z_2$-even gaugino and the brane gaugino. So, the conventional chiral superfield containing a left-handed fermion, $(Q^*,(\psi_Q)^c)$, should have an opposite $R$ charge, namely, $r$ for $Q^*$ 
and $r-1$ for $(\psi_Q)^c$.} $(Q,\psi_Q)$, 
the superfield of which has an $R$ charge $-r$, and a brane vector multiplet $(W_\mu,\Lambda)$.
Then, by employing the Noether method for the local SUSY, we find that the supersymmetric action for the bulk-brane system (up to four fermion terms) is composed of the original bulk action (\ref{bulkaction}) with the field strength tensors $G_{MNP}$ and $F_{MN}$ being replaced by the modified ones ${\hat G}_{MNP}$ and ${\hat F}_{MN}$, respectively, and the brane action as follows,
\be
{\cal L} = {\cal L}_{\rm bulk}(G\rightarrow {\hat  G},F\rightarrow {\hat F})+\delta^2(y){\cal L}_{\rm brane} \label{susyaction}
\ee 
with
\bea
{\cal L}_{\rm brane}&=& e_4 \bigg[e^{\frac{1}{2}\phi}\Big( -(D^\mu Q)^\dagger D_\mu Q+\frac{1}{2}{\bar\psi}_Q\gamma^\mu D_\mu \psi_Q +{\rm h.c.}\Big) \nonumber \\
&&\quad +\sqrt{2}irg e^{\frac{1}{4}\phi}{\bar\psi}_Q\lambda_+ Q+{\rm h.c.} -4rg^2|Q|^2-T \nonumber \\
&&\quad +e^{\frac{1}{2}\phi}\Big(\frac{1}{2}{\bar\psi}_{\mu+}\gamma^\nu\gamma^\mu \psi_Q (D_\nu Q)^\dagger +\frac{1}{2}{\bar\psi}_Q\gamma^\mu\chi_+ D_\mu Q +{\rm h.c.}\Big) \nonumber \\
&&\quad-\frac{1}{4}W_{\mu\nu}W^{\mu\nu}
+\frac{1}{2}{\bar\Lambda}\gamma^\mu D_\mu \Lambda+{\rm h.c.} \nonumber \\
&&\quad -ie\sqrt{2} e^{\frac{1}{2}\phi}Q{\bar\psi}_Q\Lambda+{\rm h.c.}-\frac{1}{2}e^2|Q|^4e^\phi\nonumber \\
&&\quad -\frac{1}{4\sqrt{2}}{\bar\Lambda}\gamma^\mu\gamma^{\nu\rho}\psi_{\mu+} W_{\nu\rho}
-\frac{i}{2\sqrt{2}}e|Q|^2 e^{\frac{1}{2}\phi}{\bar\Lambda}\gamma^\mu\psi_{\mu+}+{\rm h.c.} 
\nonumber \\
&&\quad -\frac{i}{\sqrt{2}}e|Q|^2 e^{\frac{1}{2}\phi}{\bar\chi}_+\Lambda+{\rm h.c.}\bigg].\label{braneaction}
\eea
The SUSY transformations of the brane chiral multiplet are
\be
\delta Q = \frac{1}{2}{\bar\vep}_+\psi_Q, \ \ \delta\psi_Q=-\frac{1}{2}\gamma^\mu\vep_+ D_\mu Q.
\ee
On the other hand, the SUSY transformations of the brane vector multiplet are
\bea
\delta W_\mu&=&\frac{1}{2\sqrt{2}}{\bar\vep}_+\gamma_\mu\Lambda +{\rm h.c.}, \\
\delta\Lambda &=&\frac{1}{4\sqrt{2}}\gamma^{\mu\nu}\vep_+ W_{\mu\nu}+\frac{i}{2\sqrt{2}}e|Q|^2 e^{\frac{1}{2}\phi}\vep_+.
\eea
Here the brane gauge field strength is $W_{\mu\nu}=\partial_\mu W_\nu-\partial_\nu W_\mu$ and the covariant derivatives of the brane multiplets are
\bea
D_\mu Q&=&(\partial_\mu+irgA_\mu-ieW_\mu)Q, \\
D_\mu\psi_Q&=&(\partial_\mu+i(r-1)gA_\mu-ieW_\mu+\frac{1}{4}\omega_{\mu\alpha\beta}\gamma^{\alpha\beta})\psi_Q, \\
D_\mu\Lambda &=&(\partial_\mu-igA_\mu+\frac{1}{4}\omega_{\mu\alpha\beta}\gamma^{\alpha\beta})\Lambda.
\eea
We note that the $R$ charges of the component fields in the brane chiral multiplet are different by $+1$ as known to be the case in 4D local SUSY\cite{u1rgauge}. 
The gaugino of a brane vector multiplet also has the same $R$ charge $+1$ as the bulk gravitino. 

The modified field strength tensors are
\bea
{\hat G}_{\mu m n}&=&G_{\mu mn}+\Big(J_\mu -\xi A_\mu\Big)\epsilon_{mn}\frac{\delta^2(y)}{e_2}, \label{mkr1}\\
{\hat G}_{\tau\rho\sigma}&=&G_{\tau\rho\sigma}+J_{\tau\rho\sigma}\frac{\delta^2(y)}{e_2}, \label{mkr2}\\
{\hat F}_{mn}&=&F_{mn}-(rg |Q|^2+\xi)\epsilon_{mn}\frac{\delta^2(y)}{e_2}\label{mgauge}
\eea
where $\xi=\frac{T}{4g}$ is the localized FI term, $\epsilon_{mn}$ is the 2D volume form and
\bea
J_\mu &=& \frac{1}{2}i\Big[Q^\dagger D_\mu Q-(D_\mu Q)^\dagger Q +\frac{1}{2}{\bar\psi}_Q\gamma_\mu\psi_Q-\frac{1}{2}e^{-\frac{1}{2}\phi}{\bar\Lambda}\gamma_\mu\Lambda\Big], \\
J_{\tau\rho\sigma}&=& -\frac{1}{4}{\bar\psi}_Q\gamma_{\tau\rho\sigma}\psi_Q
-\frac{1}{8}e^{-\frac{1}{2}\phi}{\bar\Lambda}\gamma_{\tau\rho\sigma}\Lambda.
\eea
Here in order to cancel the variation of the brane tension action\cite{susysolution}, we needed to modify the gauge field strength with the localized FI term\footnote{See Ref.~\cite{fi5d} for discussion on the FI term in 5D gauged supergravity.} proportional to the brane tension.  
Moreover, the modified field strength for the KR field contains a gauge non-invariant piece proportional to the localized FI term so the gauge transformation of the KR field needs to be modified to
\be
\delta_{ \Lambda} B_{mn} = \Lambda \Big(-\frac{1}{2}F_{mn}+\xi\epsilon_{mn}\frac{\delta^2(y)}{e_2}\Big).
\ee
On the other hand, the SUSY transformations of the bulk fields are the same as eqs.~(\ref{susyt1})-(\ref{susyt7}) with
$G_{MNP}$ and $F_{MN}$ being replaced by ${\hat G}_{MNP}$ and ${\hat F}_{MN}$, respectively, and
the gauge field $A_M$ being kept the same as in the no-brane case, with an exception that
the SUSY transformation of the extra components of the KR field has an additional term as
\be
\delta B_{mn}=\frac{1}{4}i {\bar\psi}_Q\vep_+ Q \epsilon_{mn}\frac{\delta^2(y)}{e_2}+{\rm h.c.}.
\ee
Furthermore, for the modified field strength tensors, we obtain the Bianchi identities as follows,
\bea
\partial_{[\mu} {\hat G}_{\nu m n]}&=&\frac{3}{4}{\hat F}_{[\mu\nu}{\hat F}_{mn]}+\bigg[\frac{i}{2}(D_{[\mu}Q)^\dagger (D_{\nu ]}Q)+\frac{1}{4}e|Q|^2W_{\mu\nu}\bigg]\epsilon_{mn}\frac{\delta^2(y)}{e_2}, \label{mbi1}\\
\partial_{[\mu} {\hat F}_{mn]}&=&-\frac{1}{3}rg \partial_\mu |Q|^2 \epsilon_{mn}\frac{\delta^2(y)}{e_2}.
\label{mbi2}
\eea
Then, by using eq.~(\ref{bulkvar}) with the modified Bianchi identities (\ref{mbi1}) and (\ref{mbi2}), we are able to cancel all the remaining variations of the brane action given in eq.~(\ref{braneaction}).

We can extend the result to the more general case with multiple branes.
When all the branes preserve the same 4D ${\cal N}=1$ SUSY, we only have to replace
the delta terms appearing in the action and the SUSY/gauge transformations: $T\delta^2(y)$ with $\sum_i T_i\delta^2(y-y_i)$,
and $f(Q)\delta^2(y)$ with $\sum_i f(Q_i)\delta^2(y-y_i)$.

\subsection{The brane-localized gravitino mass}

We introduce a gravitino mass term on the brane.
Then, the brane action is supplemented by the supersymmetric gravitino mass terms as
\be
{\cal L}_{\rm gmass}=-e_4\frac{1}{2}W_0 e^{\frac{1}{2}\psi}
({\bar\psi}_{\mu+}\gamma^{\mu\nu}C{\bar\psi}^T_{\nu+}+{\bar\psi}_1\gamma^\mu C{\bar\psi}^T_{\mu+}
+{\bar\psi}_2\gamma^\mu C{\bar\psi}^T_{\mu+}+{\bar\lambda}_+C{\bar\lambda}^T_+)+{\rm h.c.}
\ee
where $W_0$ is a constant parameter and
\be
\psi_1=\psi_{5+}+i\psi_{6+}, \quad \psi_2=\psi_{5+}-i\psi_{6+}.
\ee
We also need to modify the SUSY transformations of the extra components of the gravitino as follows,
\bea
\delta\psi_+&=&W_0 e^{\frac{1}{2}\psi}C{\bar\vep}^T_+ \frac{\delta^2(y)}{e_2}, \label{gravitino1}\\
\delta\psi_-&=&-W_0 e^{\frac{1}{2}\psi}C{\bar\vep}^T_+ \frac{\delta^2(y)}{e_2}.\label{gravitino2}
\eea
Here $e^\psi$ is the volume modulus of the extra dimensions.
Thus, similarly to the ungauged supergravity\cite{fll}, the brane gravitino mass has a nontrivial coupling to the volume modulus of the extra dimensions. Under the modified gravitino variations (\ref{gravitino1}) and (\ref{gravitino2}), the variation of the bulk gravitino linear terms would have induced singular terms for a nonzero background gauge flux. So, in order to cancel them, we needed to introduce the brane-localized gaugino mass, which is the same as the gravitino mass. 
When the superpotential depends on the brane chiral multiplets, we can infer the form of 
the brane F-term as
\be
{\cal L}_F= - e_4 e^{\psi-\frac{1}{2}\phi}|F_Q|^2 \label{fterm}
\ee
with $F_Q=\frac{\partial W}{\partial Q}$.
Consequently, similarly to the ungauged supergravity case\cite{fll}, we show that the F-term has a nontrivial coupling to the dilaton as well as the volume modulus.

\section{The supersymmetric brane solution}

In the presence of the modification in the gauge field strength in eq.~(\ref{mgauge}), 
the general warped solution obtained in Ref.~\cite{gibbons,branesol,leelud} 
is maintained up to the modified solution for the gauge field\cite{susysolution}. On the other hand, 
when the KR field and the 4D component of the gauge field are set to zero,
the modified field strength for the KR field in eqs.~(\ref{mkr1}) and (\ref{mkr2}) 
does not affect the equations of motion.

Assuming the axial symmetry in the internal space, 
it has been found that the general warped solution with 4D Minkowski space 
takes the following form\cite{gibbons,susysolution},
\bea ds^2&=&W^2(r)
\eta_{\mu\nu}dx^\mu dx^\nu+R^2(r)\bigg(dr^2
+\lambda^2 \Theta^2(r)d\theta^2\bigg), \label{wmetric} \\
{\hat F}_{mn}&=&q  \epsilon_{mn},  \label{flux}\\
\phi&=&4\ln W ,
\eea with
\be
R={W \over f_0}, \ \  \ \Theta={r \over W^4},   
\ee
\be
W^4=\frac{f_1}{f_0}, \ \ f_0=1+\frac{r^2}{r^2_0}, \ \ \
f_1=1+\frac{r^2}{r^2_1},
\ee 
where $q$ is a  constant denoting
the magnetic flux, and the two radii $r_0$, $r_1$ are given by \be
r^2_0=\frac{1}{2g^2}, \ \ r^2_1=\frac{8}{q^2}. \ee

In the warped solution, the metric has two conical singularities,
one at $r=0$ and the other at $r=\infty$, which is at finite
proper distance from the former one. The singular terms coming from the deficit angles
$\delta_i$  at these singularities need to be compensated by brane tensions
$T_i=2\delta_i(i=1,2)$ with the following matching conditions, 
\bea
\frac{\delta_1}{2\pi}&=&1-\lambda, \label{tension1}\\
\frac{\delta_2}{2\pi}&=&1-\lambda\frac{r^2_1}{r^2_0}.\label{tension2}
\eea 
Moreover, when the brane actions are invariant under the same 4D ${\cal N}=1$, 
the modified gauge field strength has two singular terms proportional
to the brane tensions at the conical singularities,
\be
{\hat F}_{mn}=F_{mn}-\epsilon_{mn}\xi_i\frac{\delta^2(y-y_i)}{e_2} \label{fi2brane}
\ee
with $\xi_i=\frac{T_i}{4g}$ for $i=1,2$. 

We need two patches of coordinates to cover the whole bulk space.
In the patch including $r=0$, 
the solution of the only non-zero component of the gauge field is
\be
A_{\theta}=-\frac{4\lambda}{q}\bigg(\frac{1}{f_1}-1\bigg)+\frac{\xi_1}{2\pi
}. \label{gaugep1} \ee 
Likewise, the gauge potential in the patch
surrounding  $r=\infty$ is \be
A_{\theta}=-\frac{4\lambda}{q}\frac{1}{f_1}-\frac{\xi_2}{2\pi
}.\label{gaugep2} \ee Hence, after connecting the gauge field
solutions in two patches by a gauge transformation and requiring
that it is single valued under $2\pi$ rotations,  we find the
following  quantization condition should hold \be \frac{4\lambda
g}{q}=n-\frac{g}{2\pi}(\xi_1+\xi_2), \ \ n \in {\mathbf Z}.
\label{quantcond} \ee In other words, we find that the FI terms
fix the Wilson line phases of the gauge potential to be
non-vanishing on the branes and can contribute to the quantization
condition for $\xi_1+\xi_2\neq 0$, \ie when $T_1+T_2\neq 0$.
This result can be regarded as a generalization of the 6D global SUSY case discussed in Ref.~\cite{lnz} to the case with nonzero bulk gauge flux.
Using the flux quantization (\ref{quantcond})
with eqs.~(\ref{tension1}) and (\ref{tension2}), we obtain the
brane tensions are related as \be
\Big(1-\frac{T_1}{4\pi}\Big)\Big(1-\frac{T_2}{4\pi}\Big)=\Big[n-
\frac{g}{2\pi}(\xi_1+\xi_2)\Big]^2.
\ee

In particular, in the  unwarped limit, \ie for $r_0=r_1$, 
the solution becomes the football solution with two equal brane
tensions $T_1=T_2=4\pi (1-\lambda)$. 
Then, since $q=4g$ and $\xi_1=\xi_2=\frac{\pi}{g}(1-\lambda)$, 
the quantization condition (\ref{quantcond}) is satisfied for $n=1$
and arbitrary $\lambda$. Thus, for $0<\lambda<1$, brane tensions can be arbitrary and positive.
This is a remarkable result, as compared to the non-SUSY brane tension action\cite{football} where 
the brane tensions are always negative because $\lambda$ is quantized as a natural number.

Moreover, unlike the general warped solution which breaks SUSY completely,
it has been shown\cite{susysolution} that for any $\lambda$, 
the football solution preserves 4D ${\cal N}=1$ SUSY.
For completeness, 
we add a brief discussion on this result given in Ref.~\cite{susysolution} in order. 
In the patch surrounding the brane at
$r=0$, the nontrivial fermionic SUSY transformations are \bea
\delta\lambda &=& i\sqrt{2} g (\gamma^5-1)\vep, \\
\delta\psi_\theta &=& \bigg[\partial_\theta+\frac{i}{2}\bigg\{1+\lambda\Big(1-\frac{2}{f_0}\Big)\bigg\}\gamma^5
+i\lambda\Big(\frac{1}{f_0}-1\Big)-i\frac{g\xi_0}{2\pi}\bigg]\vep \nonumber \\
&=&\bigg[\partial_\theta+\frac{i}{2}\bigg\{1+\lambda\Big(1-\frac{2}{f_0}\Big)\bigg\}(\gamma^5-1)\bigg]\varepsilon,
\eea where use is made of $g\xi_0=\frac{1}{4}T_0=\pi(1-\lambda)$ from
eq.~(\ref{tension1}) in the last line. Then, for a non-zero
left-handed variation parameter $\vep_+$, for which the
gaugino variation is manifestly zero, the remaining nonzero
gravitino variation is $\delta\psi_{\theta
+}=\partial_\theta\vep_+$. So, for any $\lambda$, i.e. any brane tension,
there exists a constant 4D left-handed Killing spinor $\vep_+$.
Therefore, we can see that the modified gauge potential is crucial for maintaining the 4D ${\cal N}=1$
SUSY even in the presence of arbitrary brane tensions.
It has been also shown that there appears a single chiral massless mode
of gravitino for positive brane tensions on the football\cite{susysolution}, while there are multiple massless modes of gravitino possible in the case with the non-SUSY brane tension action\cite{gravitino}.

\section{The 4D effective action}

We consider the low energy effective action containing light bulk and brane fields
after the extra dimensions are compactified on a football.
We show that the resulting 4D effective action can be described by a 4D supergravity
with the specific form of the K\"ahler potential containing a constant $U(1)_R$ FI term, 
the gauge kinetic functions and the effective superpotential from the branes.
We discuss on the effect of the bulk Green-Schwarz term on the 4D anomalies.

\subsection{The tree-level effective action}

To make a KK dimensional reduction to 4D for the supersymmetric brane solution, 
we take the separable ansatz for the 6D metric\footnote{We set the shape modulus to zero, i.e. $\xi=\psi$ in the general metric ansatz in the appendix B, because the shape modulus is massive\cite{scalarpert}.} as
\bea
ds^2=e^{-\psi(x)}g_{\mu\nu}(x)dx^\mu dx^\nu+e^{\psi(x)} ds^2_2 
\eea
where $ds^2_2$ is the 2D metric of the football solution and $\psi$ is the volume modulus.

By integrating the supersymmetric bulk-brane system (\ref{susyaction}) over the extra dimensions
with eqs.~(\ref{effg}) and (\ref{efff}), the 4D effective action for the bosonic fields apart from the brane F- and D-terms can be obtained as
\bea
{\cal L}_{\rm boson}&=&\frac{1}{2}M^4_*e_4\int d^2y e_2\bigg[R(g)-(\partial_\mu \psi)^2
-\frac{1}{4}(\partial_\mu f)^2-\frac{1}{2}e^{\psi+\frac{1}{2}f}M^{-4}_*F_{\mu\nu}F^{\mu\nu} 
-\frac{1}{2}e^{2\psi+f}(G_{\mu\nu\rho})^2
\nonumber \\
&&\qquad\qquad-\frac{1}{2}e^{-2\psi+f}\Big(\partial_\mu b-\frac{2q}{M^4_*} A_\mu-\frac{i}{M^2_P}(Q^\dagger D_\mu Q-(D_\mu Q)^\dagger Q)\Big)^2 \nonumber \\
&&\qquad\qquad-M^{-4}_*e^{-3\psi+\frac{1}{2}f}\Big(q-\frac{rg}{V}|Q|^2\Big)^2
-4g^2M^4_*(-2e^{-2\psi}+e^{-\psi-\frac{1}{2}f})\bigg] \nonumber \\
&&\quad +e_4\bigg[ -e^{-\psi+\frac{1}{2}f} (D^\mu Q)^\dagger (D_\mu Q)-\frac{1}{4}W_{\mu\nu}W^{\mu\nu}-2rg^2M^4_*e^{-2\psi}|Q|^2\bigg]\label{effaction1}
\eea
where $q=2g M^4_*$. Here we have recovered the 6D fundamental scale $M_*$ that was taken to be $M^4_*=2$. 

We redefine the scalar fields as the mixture of the dilaton and the volume modulus as
\be
s=e^{\psi+\frac{1}{2}f}, \quad t=e^{\psi-\frac{1}{2}f}.
\ee
We also dualize the 4D component field strength for the KR field as $e^f G_{\mu\nu\rho}=\epsilon_{\mu\nu\rho\tau}\partial^\tau\sigma$.
The effective action (\ref{effaction1}) then becomes
\bea
{\cal L}_{\rm boson}&=&M^2_P \sqrt{-g}\bigg[\frac{1}{2}R(g)-\frac{(\partial_\mu s)^2}{4s^2}
-\frac{(\partial_\mu t)^2}{4t^2} \nonumber \\
&&\qquad\qquad-\frac{1}{4M^2_P}sF_{\mu\nu}F^{\mu\nu}-\frac{1}{M^2_P t} (D^\mu Q)^\dagger (D_\mu Q)-\frac{1}{4M^2_P}W_{\mu\nu}W^{\mu\nu} \nonumber \\
&&\qquad\qquad-\frac{(\partial_\mu\sigma)^2}{4s^2}-\frac{1}{4t^2}\Big(\partial_\mu b-4g_R A_\mu-\frac{i}{M^2_P}(Q^\dagger D_\mu Q-(D_\mu Q)^\dagger Q)\Big)^2 \nonumber \\
&&\qquad\qquad-\frac{2g^2_RM^2_P}{s}\bigg\{1-\frac{1}{t}\Big(1-\frac{r}{2M^2_P}|Q|^2\Big)\bigg\}^2\bigg].\label{effaction2}
\eea
Here the 4D Planck scale is given by $M^2_P=M^4_* V$ with the volume of the extra dimensions $V=\lambda\pi r^2_0$, and the covariant derivative for the brane scalar is given by $D_\mu Q=(\partial_\mu+irg_RA_\mu-ieW_\mu)Q$ with the 4D effective $U(1)_R$ gauge coupling
$g_R=g/\sqrt{V}$. Here we can see that the $U(1)_R$ gauge boson gets a nonzero mass, $M^2_{A}=8g^2_RM^2_P$, 
by a Green-Schwarz mechanism due to the nonzero gauge flux.

We now find that the bosonic effective action (\ref{effaction2}) corresponds to
the 4D supergravity action with the K\"ahler potential\footnote{For simplicity, we omitted the coupling of the brane vector multiplet to the brane chiral multiplet. It can be implemented by replacing $e^{-2rg_RV_R}$ in eq.~(\ref{kahler}) with $e^{-2rg_RV_R+2eV_W}$.} $K$ containing a constant
$U(1)_R$ FI term, 
the gauge kinetic functions for the bulk and brane vector multiplets, $f_R$ and $f_W$:
\bea
K&=&-\ln\Big(\frac{1}{2}(S+S^\dagger)\Big)-\ln\Big(\frac{1}{2}(T+T^\dagger-\delta_{GS} V_R)-\frac{1}{M^2_P}{\tilde Q}^\dagger e^{-2rg_R V_R}{\tilde Q}\Big)-\frac{2\xi_R}{M^2_P} V_R, \label{kahler}\\
f_R&=& S, \qquad\qquad  f_W = 1
\eea 
where the scalar components of the moduli supermultiplets are given by
\be
S=s+i\sigma, \quad \quad T=t+\frac{1}{M^2_P}|Q|^2+ib.
\ee
Here the Green-Schwarz parameter is $\delta_{GS}=8g_R$ and the constant FI term coefficient is related to the Green-Schwarz parameter by $\xi_R=\frac{1}{4}\delta_{GS} M^2_P$.
We note that the chiral superfield $\tilde Q$ contains the component fields, $(Q^*,\psi^c_Q)$.
The nonzero scalar potential in the effective action (\ref{effaction2}) corresponds to the $U(1)_R$ D-term in the 4D effective supergravity.
We can also see that the $U(1)_R$ gauge kinetic function $f_R$ has a nontrivial $S$ modulus dependence
while the gauge kinetic function $f_W$ for the brane vector multiplet is trivial.

After making the gravitino kinetic term canonical in the KK reduction to 4D,
the effective 4D gravitino mass becomes
\be
{\cal L}_{\rm gmass}=-e_4 \frac{1}{2} W_0 e^{-\psi}{\bar\psi}_{\mu+}\gamma^{\mu\nu}C{\bar\psi}^T_{\nu+}+{\rm h.c.}.
\ee
Compared to the gravitino mass in 4D supergravity,
${\cal L}_m=-e_4\frac{1}{2}e^{K/2}W{\bar\psi}_{\mu+}\gamma^{\mu\nu}C{\bar\psi}^T_{\nu+}+{\rm h.c.}$, the effective superpotential is independent of the moduli:
\be
W=W_0.
\ee
We note that $U(1)_R$ gauge invariance requires the superpotential to take an R charge $+2$
as the gravitino $\psi_{\mu +}$ has an $R$ charge $+1$. 

We can easily generalize the results to the case when the brane matter is present at the other brane.
The 4D effective supergravity is then described by
\bea
K&=&-\ln\Big(\frac{1}{2}(S+S^\dagger)\Big) \nonumber \\
&&-\ln\Big(\frac{1}{2}(T+T^\dagger-\delta_{GS} V_R)-\frac{1}{M^2_P}\sum_{i=1,2} {\tilde Q}^\dagger_i e^{-2r_ig_R V_R}{\tilde Q}_i\Big)-\frac{2\xi_R}{M^2_P} V_R, \\
W&=& \sum_{i=1,2} W_i({\tilde Q}_i) \label{kahlergen}
\eea
where the scalar component of the $T$ modulus is generalized to
\be
T=t+\frac{1}{M^2_P}\sum_{i=1,2}|Q_i|^2+ib.
\ee
Therefore, the scalar potential takes a more general form as
\be
V_0=\frac{2g^2_RM^4_P}{s}\bigg[1-\frac{1}{t}\Big\{1-\frac{1}{2M^2_P}\Big(\sum_{i=1,2}r_i|Q_i|^2\Big)\Big\}\bigg]^2.
\ee
For a more general K\"ahler potential for the brane chiral multiplet 
$\Omega_i({\tilde Q}^\dagger_i e^{-2rg_R V_R}{\tilde Q}_i)$, we have to replace the scalar component of the $T$ modulus and the K\"ahler potential, respectively, by
\be
T=t+\sum_{i=1,2}\Omega_i(|Q_i|^2)+ib
\ee
and
\bea
K&=&-\ln\Big(\frac{1}{2}(S+S^\dagger)\Big) \nonumber \\
&&-\ln\Big(\frac{1}{2}(T+T^\dagger-\delta_{GS} V_R)-\sum_{i=1,2}\Omega_i({\tilde Q}^\dagger_i e^{-2r_ig_R V_R}{\tilde Q}_i)\Big)-\frac{2\xi_R}{M^2_P} V_R.
\eea

\subsection{The $U(1)_R$ gauge transformations}

The brane chiral multiplet $\tilde Q$ having an $R$ charge $r$ transforms under the $U(1)_R$ 
with parameter $\Phi$ (where ${\rm Re}\,\Phi|_{\theta={\bar\theta}=0}=\Lambda_R$) as 
\be
{\tilde Q}\rightarrow e^{irg_R\Phi} {\tilde Q}
\ee
while the $U(1)_R$ vector multiplet transforms as 
\be
V_R\rightarrow V_R+\frac{i}{2}(\Phi-\Phi^\dagger).
\ee
Gauge invariance of the $T$-dependent piece of the K\"ahler potential (\ref{kahler}) 
requires that, under the $U(1)_R$ gauge transformation, 
the $T$ modulus transforms nonlinearly as
\be
T\rightarrow T+\frac{i}{2}\delta_{GS} \Phi.
\ee 
This results in a shift of the axion field, $b\rightarrow b+\frac{1}{2}\delta_{GS} \Lambda_R=b+4g_R \Lambda_R$.
This is consistent with the globally well-defined KR field\cite{moduli}, given in eq.~(\ref{gwellkr}), ${\cal B}_{mn}=-b\epsilon_{mn}$, that has a gauge transformation, $\delta {\cal B}_{mn}=-\Lambda\langle {\hat F}_{mn}\rangle=-2qM^{-4}_*\Lambda\epsilon_{mn}$ with $\Lambda=\Lambda_R/\sqrt{V}$. 
On the other hand, the $S$ modulus does not transform under the $U(1)_R$.

The constant $U(1)_R$ FI term appearing in the K\"ahler potential (\ref{kahler}) causes the K\"ahler
potential to transform under the $U(1)_R$ as
\be
K\rightarrow K-i\frac{\xi_R}{M^2_P}(\Phi-\Phi^\dagger).\label{kahlervar}
\ee
So, the K\"ahler potential transforms exactly like an abelian vector superfield.
In the global SUSY case, such a variation of the K\"ahler potential would maintain the invariance of the action after the superspace integral.
In the local SUSY case, however, the action would not be invariant unless the Weyl rescaling invariance of the supergravity is supersymmetric. 
In the Weyl compensator formalism\cite{kapl}, the super-Weyl symmetry is manifest due to the chiral compensator superfield. In this case, the 4D supergravity action is written in the superspace form as
\be
S=\int d^4x\bigg[d^4\theta \,{\bf E}(-3C^\dagger C\, e^{-K/3})+\Big(\int d^2\theta\, {\cal E}C^3 W+{\rm h.c.}\Big)\bigg]\label{superact}
\ee
where $\bf E$ is the full superspace measure, $\cal E$ is the chiral superspace
measure and $C$ is the chiral compensator superfield. 
The above action (\ref{superact}) is super-Weyl invariant under the following transformations,
\bea
{\bf E}&\rightarrow& e^{2\tau+2{\bar\tau}}\, {\bf E}, \quad \quad
{\cal E}\rightarrow e^{6\tau}{\cal E}, \\
C&\rightarrow& e^{-2\tau} C, \qquad\quad
W\rightarrow W
\eea
with a complex parameter $\tau$.
On the other hand, we can show that together with eq.~(\ref{kahlervar}), the action is
also invariant under the $U(1)_R$ transformations,
\bea
C&\rightarrow& e^{-i\frac{\xi_R}{3M^2_P} \Phi}C, \\
W&\rightarrow& e^{i\frac{\xi_R}{M^2_P}\Phi} W.
\eea
Thus, since $e^{i\xi_R\Phi/M^2_P}=e^{2ig_R\Phi}$,
the effective superpotential takes an $R$ charge $+2$, as well known in the 4D supergravity with the gauged $U(1)_R$\cite{u1rgauge,binetruy}. 
When we choose the super-Weyl gauge $C=1+\theta^2 F_C$ for manifest SUSY and holomorphicity,
combining the $U(1)_R$ transformation and a super-Weyl transformation with $\tau=-i\frac{\xi_R}{6M^2_P}\Phi$ maintains the super-Weyl gauge while making the superspace action (\ref{superact}) gauge invariant. Then, the accompanying super-Weyl transform of the
superspace measure $\bf E$ and $\cal E$ means that the gravitino transforms under the $U(1)_R$.

\subsection{The bulk Green-Schwarz term and the 4D anomalies}

In order to cancel the reducible bulk anomalies, it is necessary to introduce a Green-Schwarz(GS) term\cite{greenschwarz,erler,GSterm} as follows,
\be
{\cal L}_{GS}=-kv B\wedge\Big({\rm tr}(R\wedge R)-{\tilde v}F\wedge F\Big)
\ee
with the extended gauge transformation of the KR field, $\delta B=-\frac{1}{2}\Lambda F+\omega^1_L/v$, 
where $\delta\omega^1_L=d\omega_L$ with $\omega_L$ being the gravitational 
Chern-Simons(CS) form\footnote{Since the GS term and the necessary addition of the gravitational CS term are not invariant under the SUSY transformations, we would need to add more terms in the bulk action. However, considering the complete SUSY action with them is beyond the scope of the paper.}
satisfying $d\omega_L={\rm tr}R\wedge R$. Gauge invariance would require to modify
the field strength as $G=dB+\frac{1}{2}F\wedge A-\omega_L/v$. 
Here $k,v,{\tilde v}$ are calculable for the given bulk fermion content.

From eq.~(\ref{gwellkr}), we can rewrite the KR field in terms of the globally well-defined one as $B={\cal B}+\frac{1}{2}\langle A\rangle\wedge {\cal A}$.
Then, under the gauge transformations, $\delta_\Lambda{\cal B}=-\Lambda\langle {\hat F}\rangle$
and $\delta_\Lambda {\cal A}=d\Lambda$, 
the bulk Green-Schwarz term transforms
\be
\delta_\Lambda {\cal L}_{GS}=- kv\Big(-\Lambda \langle{\hat F}\rangle+\frac{1}{2}\langle A\rangle\wedge d\Lambda\Big)
\wedge \Big({\rm tr}(R\wedge R)-{\tilde v}F\wedge F\Big). \label{gsterm}
\ee
When we focus on the gauge part, the gauge variation of the Green-Schwarz term becomes
\be
\delta_\Lambda {\cal L}_{GS}=-kv{\tilde v}\bigg[\frac{1}{2}\Lambda\langle {\hat F}\rangle\wedge F\wedge F
+\Big(\Lambda\langle A\rangle\wedge F\wedge F\Big|_{r=\infty}-\Lambda \langle A\rangle \wedge F\wedge F\Big|_{r=0}\Big)\bigg]
\ee
where use is made of eqs.~(\ref{fi2brane}), (\ref{gaugep1}) and (\ref{gaugep2}).
The first term is a bulk term giving rise to the 4D anomalies induced by the gauge flux.
It is also present in the Salam-Sezgin vacuum\cite{SS} without brane tensions where there appear 4D chiral massless modes of the bulk fermions even without orbifold projections.
On the other hand, the last two terms correspond to the variation of the effective Chern-Simons 
action\cite{csterm}.
Since $\langle A\rangle$ does not vanish at the branes due to the localized FI terms,
they generate the $U(1)_R$ gauge anomalies on the boundaries, $r=0$ and $r=\infty$. 
The $U(1)_R$-mixed gravitational anomalies
are also induced both in the bulk and on the boundaries in a similar fashion.
Since the localized anomalies are proportional to the arbitrary localized FI terms, in order to cancel the localized anomalies, we need to introduce $R$ charged fermions on the branes. 
However, since the localized anomalies are restrictive, i.e. the $U(1)_R$-mixed gravitational anomalies are proportional to the gauge anomalies depending on the bulk fermion content via ${\tilde v}$, the brane fermion content should be constrained unless there are additional localized Green-Schwarz terms\cite{lnz}.   
The $U(1)_R$ anomaly cancellation with/without the Green-Schwarz term has been discussed in 4D supergravity context\cite{u1r}.

\section{Modulus stabilization and $U(1)_R$ mediation}

Although the 4D scalar potential stabilizes one of the moduli, the $T$ modulus, due to a nonzero $U(1)_R$ D-term, the remaining $S$ modulus needs to be fixed too.
We consider a possibility of having the $S$ modulus stabilized by a $S$-dependent superpotential
due to the bulk gaugino condensates.
Then, introducing a brane F-term and/or a brane D-term as the uplifting potential,
we can have all the moduli stabilized at the Minkowski vacuum. 
Consequently, due to the shift of the $T$ modulus from the value determined only by the $U(1)_R$ D-term, we show that the brane scalar gets a nonzero soft mass which is proportional to the $R$ charge of the brane chiral multiplet. We call this kind of mechanism of generating the soft mass ``$U(1)_R$
mediation''.

\subsection{Modulus stabilization}

From the effective action (\ref{effaction2}), we can read the 4D effective scalar potential as 
\be
V_0=\frac{2g^2_RM^4_P}{s}\bigg[1-\frac{1}{t}\Big(1-\frac{r}{2M^2_P}|Q|^2\Big)\bigg]^2.
\ee
So, we find that the minimum of the potential occurs at $t=1$ and $|Q|=0$.
Then, at this minimum, the tree-level brane scalar mass vanishes. This is due to
the cancellation between the tree-level brane-localized scalar mass term 
and the flux-induced mass term. 
Moreover, we note that the vacuum energy at the minimum vanishes without SUSY breakdown. 
Since the value of the $S$ modulus is not determined, however,
there should be additional potential terms that arise in stabilizing the $S$ modulus by some other mechanism.

In order to stabilize the $S$ modulus, we assume that the bulk non-perturbative dynamics generates
a modulus potential with $S$-dependent superpotential $W(S)$.
Then, the additional contribution to the 4D scalar potential is
\bea
V_1&=&\frac{e^K}{M^2_P}\bigg[\Big|\frac{\partial W}{\partial S}+\frac{\partial K}{\partial S}W\Big|^2 K^{-1}_{SS^\dagger}-2|W|^2\bigg] \nonumber \\
&=&\frac{1}{M^2_P st}\bigg[(S+S^\dagger)^2\Big|\frac{\partial W}{\partial S}-\frac{1}{S+S^\dagger}W\Big|^2
-2|W|^2\bigg].
\eea
The extremum conditions for the total scalar potential $V=V_0+V_1$, $\partial_S V=0$, $\partial_T V=0$ and $\partial_Q V=0$, are solved approximately\footnote{Because of the $T$ dependence of the scalar potential coming from the gaugino condensates, the minimum value of $t$ is shifted from $t=1$.
Moreover, due to the $S$ dependence of the $U(1)_R$ D-term, the minimum value of $s$ is also shifted
compared to the one determined only by $V_1$.} by $t\simeq 1$, $Q=0$ and $S$ solving $F_S\simeq 0$.

For instance, we can consider double gaugino condensates\cite{gauginocondense} in an extended bulk theory with anomaly-free non-abelian gauge groups\footnote{See Ref.~\cite{anomalyfree} for some anomaly-free models containing $U(1)_R$.}. 
In this case, assuming that the $U(1)_R$ symmetry
is broken spontaneously by the VEV of the $R$ charged scalars, the superpotential takes a racetrack form\cite{racetrack}
\be
W(S)= \Lambda_1 e^{-\beta_1 S}-\Lambda_2 e^{-\beta_2 S}. \label{superw}
\ee
Here we ignored a possible modification to the gauge kinetic function 
due to the supersymmetric completion of the Green-Schwarz term (\ref{gsterm}) and
assumed that $\Lambda_1$ and $\Lambda_2$ have the $R$ charge $+2$ due to the presence of matter
fields\cite{carloscasas}. This is in contrast to the fact that the double gaugino condensates are not possible being consistent with the global $U(1)_R$ in the heterotic string\cite{u1global} where the $S$ modulus is shifted by an imaginary amount under the $U(1)_R$. 
The matter fields charged under the condensing gauge groups could
give additional contributions to the $U(1)_R$ D-term but we assumed that an extra singlet
having an opposite $R$ charge to the matter fields cancels those contributions.

Then, we find that the $F_S=0$ condition is
\be
\Lambda_1 e^{-\beta_1 S}(1+\beta_1(S+S^\dagger))=\Lambda_2 e^{-\beta_2 S}(1+\beta_2(S+S^\dagger)).
\label{fixs}
\ee
This fixes both ${\rm Re}\, S$ and ${\rm Im}\, S$.
When the beta functions of the two condensing gauge groups are similar, $|\beta_1-\beta_2|\ll \beta_1$,
the solution to eq.~(\ref{fixs}) occurs at a large ${\rm Re}\, S$ where the superpotential description with eq.~(\ref{superw}) is reliable.

\subsection{Uplifting and soft masses}

After fixing the $S$ modulus, however, the vacuum energy becomes nonzero and negative. 
Therefore, we need to uplift the vacuum energy to zero.
To this purpose, from the 4D reduction of eq.~(\ref{fterm}), 
we derive a F-term potential at the hidden brane as
\be
V_2=\frac{1}{s}|F_{Q'}|^2.
\ee 
Instead of the brane F-term, from the 4D reduction of the brane D-term in eq.~(\ref{susyaction}), 
we can consider a D-term potential at the hidden brane as the uplifting potential,
\be
V_3=\frac{1}{2t^2}D^2.
\ee 

For generalities, including both brane F- and D-terms for the uplifting potential, 
the full 4D scalar potential becomes
\bea
V&=& V_0+V_1+V_2+V_3 \nonumber \\
&=&\frac{2g^2_RM^4_P}{s}\bigg[1-\frac{1}{t}\Big(1-\frac{r}{2M^2_P}|Q|^2\Big)\bigg]^2 \nonumber \\
&&+\frac{1}{M^2_P st}\bigg[(S+S^\dagger)^2\Big|\frac{\partial W}{\partial S}-\frac{1}{S+S^\dagger}W\Big|^2
-2|W|^2\bigg]+\frac{1}{2t^2}D^2+\frac{1}{s}|F_{Q'}|^2.
\eea
Then, $\partial_Q V=0$ is satisfied for $Q=0$, which is the minimum for $r(t-1)>0$.
In the case with $Q=0$, 
the remaining extremum conditions, $\partial_S V=0$ and $\partial_T V=0$, are
\bea
0&=&-\frac{4g^2_RM^4_P}{(S+S^\dagger)^2}\Big(1-\frac{1}{t}\Big)^2 -\frac{2}{(S+S^\dagger)^2}|F_{Q'}|^2\nonumber \\
&&\quad +\frac{2}{M^2_P t}\frac{\partial}{\partial S}\bigg\{\frac{1}{S+S^\dagger}\Big[(S+S^\dagger)^2\Big|\frac{\partial W}{\partial s}-\frac{1}{S+S^\dagger}W\Big|^2-2|W|^2\Big]\bigg\}, \label{sderv}\\
0&=&\frac{4g^2_R M^4_P}{st^2}\Big(1-\frac{1}{t}\Big)-\frac{1}{t^3}D^2
-\frac{1}{M^2_Pst^2}\bigg[(S+S^\dagger)^2\Big|\frac{\partial W}{\partial S}-\frac{1}{S+S^\dagger}W\Big|^2-2|W|^2\bigg].\label{tderv}
\eea
Compared to the case without the uplifting potential, from eq.~(\ref{tderv}), the $T$ modulus is shifted to 
\be
t=\frac{1+\frac{1}{2}\alpha D^2}{1-\frac{1}{2}\alpha tV_1}
\ee
where $\alpha\equiv\frac{s}{2g^2_R M^4_P}$ and $V_1$ is the scalar potential coming from the gaugino
condensates. Thus, we note that the F-term at the hidden brane does not contribute directly to the shift of the $T$ modulus because it is independent of the $T$ modulus.
On the other hand, the $S$ modulus is also determined from eq.~(\ref{sderv}) for the fixed $t$.
We take the vacuum energy to be zero by choosing the brane D-term as
\be
D^2=\frac{-2tV_1\Big(1-\frac{1}{4}\alpha tV_1\Big)-\frac{2}{s}|F_{Q'}|^2}{1+\frac{\alpha}{s}|F_{Q'}|^2}.
\ee
Consequently, after minimizing the moduli potential, we find that a nonzero mass for the brane scalar 
with nonzero $R$ charge is generated and it is given by the following general formula,
\bea
m^2_Q&=& K^{-1}_{QQ^\dagger}\frac{\partial^2 V}{\partial Q\partial Q^\dagger}\Big|_{Q=0} \nonumber \\
&=&\frac{2g^2_RM^4_P}{s}(t-1)\frac{r}{tM^2_P} \nonumber \\
&=& \frac{D^2+tV_1}{1-\frac{1}{2}\alpha tV_1}\frac{\frac{1}{2}r}{tM^2_P}.\label{gscalarmass}
\eea
We note that the brane scalar mass does not depend on the $U(1)_R$
gauge coupling. This is because the effective interaction between the visible and hidden sectors is suppressed by the mass squared of the $U(1)_R$ vector multiplet, $M^2_R=8g^2_R M^2_P$. 
That is, the gauge coupling dependence is cancelled out by the one of the suppression scale for the $U(1)_R$ mediation. 
It would be interesting to compare the $U(1)_R$ mediation to some relevant results in the literature on the SUSY mediation via an additional anomalous or non-anomalous $U(1)$ gauge field\cite{u1prime,dudas,choi}.

First we consider the case with the brane D-term domination by setting $F_{Q'}=0$.
Assuming $\alpha t|V_1|\ll 1$ and ignoring
the vacuum energy contribution due to a nonzero F-term for the $S$ modulus,
$D^2\simeq -2tV_1$ and $V_1\simeq -\frac{2|W|^2}{M^2_Pst}$ so we require the brane D-term to be $D^2\simeq \frac{4|W|^2}{M^2_Ps}$.
Then, from eq.~(\ref{gscalarmass}) with $F_{Q'}=0$, the brane scalar mass becomes
\be
m^2_Q\simeq \frac{r}{st}\frac{|W|^2}{M^4_P}=r \,m^2_{3/2}
\ee
where use is made of the gravitino mass given by
$m^2_{3/2}=e^{K}\frac{|W|^2}{M^4_P}=\frac{1}{st}\frac{|W|^2}{M^4_P}$.
For the positive brane scalar mass squared, we require the $R$ charge of the brane scalar to be positive. 
After integrating out the $U(1)_R$ vector multiplet, the effective operator for generating the soft mass would be written as
$\int d^4\theta \frac{g^2_R}{M^2_R}W'_\alpha W^{'\alpha}{\tilde Q}^\dagger {\tilde Q}$
with $W'_\alpha$ being the superfield strength for the hidden sector gauge field.
We note, however, that from all the known microscopic models for generating the D-term uplifting potential\cite{dudas,choi,dtermlift},
such a large D-term generically gives rise to a very heavy gravitino.

Secondly we take the case with the brane F-term domination for which $D=0$.
Then, similarly to the D-term domination case, for $\alpha r|V_1|\ll 1$,
we need the brane F-term to be $|F_{Q'}|^2\simeq -4stV_1\simeq \frac{8|W|^2}{M^2_P}$.
Thus, from eq.~(\ref{gscalarmass}) with $D=0$, the brane scalar mass becomes
\be
m^2_Q\simeq -\frac{r}{st}\frac{|W|^2}{M^4_P}=-r \,m^2_{3/2}.
\ee
Thus, the brane scalar mass gets an opposite sign compared to the brane D-term
domination, so the $R$ charge of the brane scalar must be negative for the positive
scalar mass squared. In the F-term domination,
the effective operator for generating the soft mass would be written as
$\int d^4\theta \frac{g^2_R}{M^2_R}{Q{'^\dagger} Q'}{\tilde Q}^\dagger {\tilde Q}$
with $Q'$ being the hidden sector chiral superfield\footnote{Even in the case that the hidden chiral superfield does not have an $R$ charge, it has a gauge coupling to the $U(1)_R$ vector multiplet as seen from the K\"ahler potential (\ref{kahlergen}).}.

Therefore, for either brane D- or F-term domination,
the tree-level soft mass due to the $U(1)_R$ mediation can be positive for the appropriate 
$R$ charge assignment so that it dominates over the anomaly mediation.
Particularly, when the $R$ charges of sleptons are nonzero, we can cure the problem
of the negative slepton masses in the anomaly mediation\cite{anomalymed}. 

Before ending the section, we also make a remark on the gaugino masses for the Standard Model gauge
group. When the SM gauge fields are localized on the brane, there is no gaugino mass at the tree level
because the brane gauge kinetic term is trivial, i.e. $f_W=1$.
Thus, one can argue that the gaugino masses are generated at one-loop due to the anomaly 
mediation\cite{anomalymed}.
In this case, the gaugino masses are suppressed by the loop factor, compared to the gravitino mass or the scalar mass.

\section{Conclusion}

We have constructed a consistent SUSY action for brane matter multiplets
in a 6D chiral gauged supergravity. Introducing brane chiral multiplets charged under the $U(1)_R$,
we derived the supersymmetric $U(1)_R$ coupling to the brane by modifying both the gauge field strength
and the field strength for the KR field together with the necessary modifications of the fermionic
SUSY transformations. We also notify that the modified field strength for the KR field is consistent with SUSY and $U(1)_R$ symmetry only at the expense of modifying the SUSY and gauge transformations of the KR field with the singular terms, respectively. 

The singular modification of the field strength does not challenge the attempt of finding the meaningful result but rather is necessary for obtaining the consistent 4D effective action. 
We showed that after a dimensional reduction to 4D on the SUSY football background,
the obtained low energy effective action with light bulk and brane modes can be consistently 
reproduced by the corresponding 4D effective supergravity containing the $U(1)_R$ gauge symmetry.
The resulting $U(1)_R$ gauge symmetry in the 4D effective theory is anomalous as the bulk Green-Schwarz term
generates the 4D anomalies in the bulk and on the boundaries.
Moreover, the effective scalar potential coming from the $U(1)_R$ D-term stabilizes
one of two moduli, i.e. the $T$ modulus, at the SUSY Minkowski vacuum.
Due to the Green-Schwarz mechanism with the gauge flux, the $U(1)_R$ gauge symmetry is spontaneously broken by eating up the axionic scalar partner of the $T$ modulus, so the mass of the $U(1)_R$ gauge field is of order the 4D Planck scale up to the $U(1)_R$ gauge coupling.
A brane scalar with nonzero $R$ charge appears in the $U(1)_R$ D-term and the mass of the
brane scalar vanishes at the minimum of the $T$ modulus due to the cancellation between the
$R$-charge dependent brane-localized mass term and the flux-induced mass term.
However, the $S$ modulus needs to be stabilized for avoiding an unacceptable effect on the equivalence principle and the cosmology.

For the stabilization of the $S$ modulus, 
we consider the bulk gaugino condensates in an extended bulk theory with the
product of the $U(1)_R$ and non-abelian gauge groups.  
Then, since the bulk gauge kinetic function depends linearly on the $S$ modulus,
the effective superpotential for the double gaugino condensates takes a racetrack form for the $S$ modulus.
In the process of the $S$ modulus stabilization, we introduce brane F- and/or D-term potentials
to uplift a negative vacuum energy to zero.
Because the scalar potential coming from the gaugino condensates or the brane D-term
depends on the $T$ modulus, after the $S$ modulus stabilization, the minimum value
of the $T$ modulus is shifted from $t=1$ that would have been obtained only for the $U(1)_R$ D-term.
Thus, we obtained a nonzero soft mass for the brane scalar proportional to its $R$ charge. 
For this reason, we owe the obtained SUSY breaking to $U(1)_R$ mediation.
For the $R$ charge of order one, the scalar soft mass is of the same order as the gravitino mass.
Depending whether brane F- or D-term dominates in the uplifting potential, 
the brane scalar soft mass squared can be positive or negative for a fixed $R$ charge.
Therefore, according to the nature of the uplifting potential coming from the hidden brane, an appropriate $R$ charge assignment for the brane scalar is required to get a positive scalar soft mass squared.

We have discussed briefly on the $U(1)_R$ anomalies only from the bulk Green-Schwarz counterterm.
Since the integrated 4D $U(1)_R$ anomalies in our case are constrained in the same way as in 4D supergravity, some important developments in the study of the anomaly-free $U(1)_R$ symmetry in 4D supergravity\cite{u1r} would be applicable to our effective 4D supergravity. 
The 6D gauged supergravity with the consistent $U(1)_R$ gauge symmetry might have a variety of applications in particle physics and cosmology, for instance, to consider
the D-term inflation\cite{binetruy} and explain the baryon/lepton number conservation\cite{baryon,u1r}, 
the $\mu$ problem\cite{muproblem} and the fermion mass hierarchy in the context
of the horizontal $U(1)$ symmetry\cite{fermionmass}. 
We leave the detailed analysis on a realistic model building for one problem or another as a future work.

\acknowledgments

The author thanks Hans Peter Nilles and Adam Falkowski for valuable comments.
He acknowledges the cosmology group at Perimeter Institute for warm hospitality and interesting discussion. The work of the author is supported by the DOE Contracts DOE-ER-40682-143 and
DEAC02-6CH03000.

\appendix

\section{Notations and Conventions}\label{sec:A}{\small

We use the metric signature $(-,+,+,+,+,+)$ for the 6D metric. The
index conventions are the following: (1) for the Einstein indices
we use  $M,N,\cdots=0,\cdots,5,6$ for the 6D indices,
$\mu,\nu,\cdots,=0,\cdots,3$ for the 4D indices and
$m,n,\cdots=5,6$ for the internal 2D indices,  (2) for the Lorentz
indices we use $A,B,\cdots=0,\cdots,5,6$ for the 6D indices,
$\alpha,\beta,\cdots=0,\cdots,3$  for the 4D indices and$a,b,\cdots=5,6$  for the internal 2D indices.

We take the gamma matrices in the locally flat
coordinates, satisfying
$\{\Gamma_A,\Gamma_B\}=2\eta_{AB}$, to be \bea
\Gamma_\alpha&=&\sigma^1\otimes\gamma_\alpha, \ \
\Gamma_5=\sigma^1\otimes\gamma_5, \ \ \Gamma_6=\sigma^2\otimes{\bf
1}, \eea where $\g$'s are the 4D gamma matrices with
$\gamma^2_5=1$ and $\sigma$'s are the  Pauli matrices with
$[\s^i,\s^j]=2i \ep_{ijk} \s^k$, with $i,j,k=1,2,3$, \be
\s^1 = \left(\begin{array}{ll}0 & 1 \\
1 & 0 \end{array}\right), \ \ \ \s^2 = \left(\begin{array}{lr}0 & -i \\
i & 0 \end{array}\right), \ \ \ \s^3 = \left(\begin{array}{lr}1 & 0 \\
0 & -1 \end{array}\right).
\ee
The convention for 4D gamma matrices is that
\be
\g^\a = \left(\begin{array}{ll}0 & \s^\a \\
\bar{\s}^\a & 0 \end{array}\right), \ \  \g^5 = \left(\begin{array}{lr} {\bf 1} & 0 \\
0 & -{\bf 1} \end{array}\right), \ee with $\s^\a=({\bf 1}, \s^i)$
and $\bar{\s}^\a=(-{\bf 1}, \s^i)$. The chirality projection
operators are defined as $P_L=(1+\gamma^5)/2$ and
$P_R=(1-\gamma^5)/2$. We note that $\G^{\a 5}={\bf 1} \otimes \g^\a \g^5$, $\G^{\a
6}= i \s^3 \otimes \g^\a$, and $\G^{56}=i \s^3 \otimes \g^5$.
In the text, we use the notation, ${\bar\chi}=-i\chi^\dagger \Gamma^0$, etc.
The curved gamma matrices on the other hand are given in terms of the ones in the locally
flat coordinates as $\G^M= e^{~M}_A \G^A$ where $e^{~M}_A$ is the 6D vielbein. 

The antisymmetrization of the gamma matrices is defined as
\be
\Gamma^{M_1 M_2\cdots M_n}=\Gamma^{[M_1}\Gamma^{M_2}\cdots \Gamma^{M_n]}=\frac{1}{n!}\sum_p (-1)^p
\Gamma^{M_1}\Gamma^{M_2}\cdots \Gamma^{M_n}
\ee
where $\sum_p$ is the summation over all permutations.
The 6D chirality operator is given by
\be
\Gamma_7=\Gamma_0\Gamma_1\cdots\Gamma_6=\sigma^3\otimes{\bf 1}.
\ee

\section{Spin connection}\label{sec:B}{\small

We include the scalar modes in the general warped solution as
\be
ds^2=e^{-\psi}W^2\eta_{\mu\nu}dx^\mu dx^\nu 
+e^{\xi}(d\rho^2+e^{2(\psi-\xi)}a^2d\theta^2).
\ee
Then, the nonzero vielbein components are given by
\bea
e^\alpha_\mu&=&e^{-\frac{1}{2}\psi}W\delta^\alpha_\mu, \\
e^a_m&=&\left(\begin{array}{ll}\cos\theta & -\sin\theta \\
\sin\theta & \cos\theta \end{array}\right)\left(\begin{array}{ll} e^{\frac{1}{2}\xi} & 0 \\ 0 &
a e^{\psi-\frac{1}{2}\xi}\end{array}\right).
\eea
Therefore, the nonzero components of the spin connection are
\bea
\omega^\alpha\,_\beta&=&\frac{1}{2}(\eta^{\alpha \rho} \eta_{\beta \mu} \partial_\rho \psi - \delta_\beta^\rho \delta^\alpha_\mu \partial_\rho \psi) dx^\mu, \\
\omega^\alpha\,_5&=&\cos\theta\bigg(\frac{W'}{W}-\frac{\psi'}{2}\bigg)W e^{-\frac{1}{2}(\psi+\xi)}
\delta^\alpha_\mu  dx^\mu  \nonumber \\
&& - \frac{1}{2}\cos\theta \eta^{\alpha  \beta}  \partial_\beta \xi W^{-1}e^{\frac{1}{2}(\psi+\xi)}d\rho \nonumber \\
&&-\sin\theta  \eta^{\alpha \beta}   \partial_\beta(\psi-\frac{1}{2}\xi)aW^{-1}e^{\frac{1}{2}(3\psi-\xi)}d\theta, \\
\omega^\alpha\,_6&=&\sin\theta\bigg(\frac{W'}{W}-\frac{\psi'}{2}\bigg)W e^{-\frac{1}{2}(\psi+\xi)}
\delta^\alpha_\mu   dx^\mu \nonumber \\
&& - \frac{1}{2} \sin\theta \eta^{\alpha \beta} \partial_\beta \xi W^{-1}e^{\frac{1}{2}(\psi+\xi)}d\rho \nonumber \\
&&+\cos\theta \eta^{\alpha \beta}  \partial_\beta (\psi-\frac{1}{2}\xi)aW^{-1}e^{\frac{1}{2}(3\psi-\xi)}d\theta, \\
\omega^5\,_6&=&\bigg[1-\Big(\psi'-\frac{1}{2}\xi'+\frac{a'}{a}\Big) a e^{\psi-\xi}\bigg]d\theta\equiv \omega d\theta
\eea
where prime denotes the derivative with respect to $\rho$.

We take the case with $\xi=\psi$ for which the shape modulus is decoupled\cite{scalarpert}.
In order to determine the modulus coupling of the brane gravitino mass in section 3.3,
the following components of the spin connection contracted with the vielbein, $\omega_{ABC}\equiv e^M_A \omega_{MBC}$, were used in the text,
\bea
\omega_{\alpha 56}&=&\omega_{5\alpha 6}=0, \\
\omega_{5\alpha 5}&=&-\frac{1}{2}\partial_\alpha \psi \,e^{\frac{1}{2}\psi} W^{-1}=\omega_{6\alpha 6}.
\eea

\section{The effective action with the delta terms}\label{sec:C}{\small

In this section, we present the details for deriving the effective action from the modified 
field strengths with the delta terms.

\subsection{The globally well-defined KR field}

We decompose the gauge field into the background value and the fluctuation as
$A=\langle A\rangle +{\cal A}$. The gauge field has different background values
in the different patches, being connected by a background gauge transformation, $\delta_{\Lambda_0}\langle A\rangle=d\Lambda_0$, while $\delta_{\Lambda_0}{\cal A}=0$.  
Likewise, the gauge field strength is also decomposed to
$F=dA=\langle F\rangle +d{\cal A}\equiv \langle F\rangle +{\cal F}$.

Under the background gauge transformation, the KR field transforms as $\delta_{\Lambda_0}B_{mn}=\Lambda_0(-\frac{1}{2}F_{mn}+\xi\epsilon_{mn}\frac{\delta^2(y)}{e_2})$, $\delta_{\Lambda_0}B_{m\mu}=-\frac{1}{2}\Lambda_0F_{m\mu}$ and $\delta_{\Lambda_0}B_{\mu\nu}=-\frac{1}{2}\Lambda_0 F_{\mu\nu}$, 
so it is not globally well-defined\footnote{Compared to Ref.~\cite{moduli}, we also have a singular delta term in the gauge transform.}. 
That is, the background gauge transformation is not of the exact form.
The gauge transformation of the derivative of $B$ is given by
\bea
\delta_{\Lambda_0}dB&=&-\frac{1}{2}d\Lambda_0 \wedge F \nonumber \\
&=&-\frac{1}{2}d\Lambda_0\wedge (\langle F\rangle+d{\cal A}) \nonumber \\
&=&-\frac{1}{2}d\Lambda_0\wedge d{\cal A}=\frac{1}{2}d(d\Lambda_0\wedge {\cal A})
=\frac{1}{2}\delta_{\Lambda_0}d(\langle A\rangle\wedge{\cal A})
\eea 
where use is made of $d\Lambda_0\wedge \langle F\rangle=0$.
So, we find that the singular term in the background gauge transform of the KR field 
does not affect the gauge transform of the derivative of $B$. 
Thus, we define the globally well-defined KR field as in the case without the delta term as
\be
{\cal B}=B-\frac{1}{2}\langle A\rangle\wedge {\cal A}. \label{gwellkr} 
\ee
Then, we can show that $\delta_{\Lambda_0}d{\cal B}=d(\delta_{\Lambda_0}{\cal B})=0$.
In this case, the redefined $\cal B$ is globally well-defined if there is a solution
for a one-form $C$ satisfying
\be
\delta{\cal B}_{mn}=\Lambda_0\Big(-\frac{1}{2} F_{mn}+\xi\epsilon_{mn}\frac{\delta^2(y)}{e_2}\Big) 
-\frac{1}{2}(d\Lambda_0\wedge {\cal A})_{mn}+(dC)_{mn}=0,
\ee
\be
\delta{\cal B}_{m\mu}=-\frac{1}{2}\Lambda_0F_{m\mu}-\frac{1}{2}(d\Lambda_0\wedge {\cal A})_{m\mu}+(dC)_{m\mu}=0, 
\ee
and 
\be
\delta{\cal B}_{\mu\nu}=-\frac{1}{2}\Lambda_0F_{\mu\nu}-\frac{1}{2}(d\Lambda_0\wedge {\cal A})_{\mu\nu}+(dC)_{\mu\nu}=0.
\ee

Now we consider the general gauge transformation of the derivative of $\cal B$ as
\be
\delta (\partial_\mu {\cal B}_{mn}+\partial_m{\cal B}_{n\mu}+\partial_n{\cal B}_{\mu m})
=\partial_\mu\Lambda \Big(-\langle F_{mn}\rangle+\xi\epsilon_{mn}\frac{\delta^2(y)}{e_2}\Big)-\frac{1}{2}\partial_\mu \Lambda (\partial_m{\cal A}_n-\partial_n{\cal A}_m).
\ee 
Thus, we obtain
\be
\delta {\cal B}_{mn}=-\Lambda \langle {\hat F}_{mn}\rangle.
\ee
Moreover, the derivative of the redefined KR field is written in terms of the gauge field background
as
\be
d{\cal B}=dB-\frac{1}{2}\langle F\rangle\wedge {\cal A}+\frac{1}{2}\langle A\rangle\wedge d{\cal A}.
\ee
Therefore, we can rewrite the decomposed field strength for the KR field as 
\bea
G&=& dB+\frac{1}{2}\langle F\rangle \wedge {\cal A}+\frac{1}{2}d{\cal A}\wedge \langle A\rangle
\nonumber \\
&=&d{\cal B}+\langle F\rangle \wedge {\cal A}.
\eea
That is, in terms of the components, it is written as
\be
G_{\mu mn}=\partial_\mu {\cal B}_{mn}+\partial_m{\cal B}_{n\mu}-\partial_n{\cal B}_{m\mu}+\langle F_{mn}\rangle {\cal A}_\mu.
\ee

\subsection{The effective matter coupling of the globally well-defined KR field}

We present the details on the effective action coming from the modified field strength for the KR field.

For the globally well-defined KR field ${\cal B}$,
the modified field strength for the KR field is given by
\bea
{\hat G}_{\mu mn}&=&\partial_\mu {\cal B}_{mn}+\partial_m{\cal B}_{n\mu}-\partial_n{\cal B}_{m\mu}+\langle F_{mn}\rangle {\cal A}_\mu+\Big(J_\mu-\xi{\cal A}_\mu\Big)\epsilon_{mn}\frac{\delta^2(y)}{e_2} \nonumber \\
&=&\partial_\mu {\cal B}_{mn}+\partial_m{\cal B}_{n\mu}-\partial_n{\cal B}_{m\mu}+\langle {\hat F}_{mn}\rangle {\cal A}_\mu +J_\mu \epsilon_{mn}\frac{\delta^2(y)}{e_2}.\label{modifyg}
\eea

Restricting ourselves to the football SUSY solution, we assume that the solutions are separable as
\bea
ds^2&=& e^{-\psi(x)}g_{\mu\nu}(x)dx^\mu dx^\nu + e^{\psi(x)} ds^2_2, \\
F_{mn}&=& q \epsilon_{mn}, \quad \phi=f(x) 
\eea
where $ds^2_2$ and $\epsilon_{mn}$ are the 2D metric and volume form for the static solution. 
Then, from the equation for the KR field, 
\be
\partial_M(\sqrt{-g}e^\phi {\hat G}^{MNP})=0,
\ee
we obtain the solution,
\be
{\hat G}_{\mu mn}=e^{3\psi-f}C_\mu \epsilon_{mn} \label{solg}
\ee
with
\be
\partial_m C_\mu=0, \quad \partial_\mu (\sqrt{-g}C^\mu)=0.
\ee
Taking the ans\"atze, ${\cal B}_{m\mu}=-\epsilon_{mn}\partial^n W_\mu$ and ${\cal B}_{mn}=-b(x)\epsilon_{mn}$, eq.~(\ref{solg}) with (\ref{modifyg}) becomes
\be
\Box^{(2)}W_\mu=-\partial_\mu b+q{\cal A}_\mu+J_\mu\frac{\delta^2(y)}{e_2}-e^{3\psi-f}C_\mu.
\ee
Due to the Stokes theorem for compact dimensions, 
integration of the left-hand side over the extra dimensions must vanish.
Thus, we obtain
\be
e^{3\psi-f}C_\mu = -\partial_\mu b+q {\cal A}_\mu +\frac{J_\mu}{V}
\ee
where the volume of the extra dimensions $V=\int d^2y e_2$.
Therefore, after integrating out the 4D vector component of
the globally well-defined KR field, 
the kinetic term for the KR field with the modified field strength becomes
\bea
{\cal L}_{KR}&=&-\int d^2y e_6\frac{1}{4}e^\phi{\hat G}_{\mu mn}{\hat G}^{\mu mn} \nonumber \\
&=&- e_4 \frac{V}{2}e^{-2\psi+f}\Big(\partial_\mu b-q{\cal A}_\mu -\frac{J_\mu}{V}\Big)^2.\label{effg}
\eea
This result is used in the text in section 5.1.
We can see that there appears a coupling of the gauge boson to the axion, which is proportional to the flux $q$. This is nothing but a spontaneous breakdown of the $U(1)_R$ gauge theory 
by Green-Schwarz mechanism.

\subsection{The effective action for the $U(1)_R$ D-term for a brane scalar}

Next we also consider the details on the effective action coming from the modified gauge field
strength.

We first consider the Bianchi identity (\ref{mbi2}) for the modified gauge field strength as
\be
\partial_\mu {\hat F}_{mn}+\partial_m F_{n\mu}-\partial_n F_{m\mu}=-rg\partial_\mu |Q|^2\frac{\delta^2(y)}{e_2}
\epsilon_{mn}.
\ee
After decomposing the field strength into the background value and the fluctuation,
the modified field strength is given by
\be
{\hat F}_{mn}=\langle {\hat F}_{mn}\rangle +{\cal F}_{mn}-rg|Q|^2\frac{\delta^2(y)}{e_2}\epsilon_{mn}.
\label{decompfmn}
\ee
Inserting the above expression into the Bianchi identity, we get the Bianchi identity for
the fluctuation as
\be
\partial_\mu {\cal F}_{mn}+\partial_m{\cal F}_{n\mu}-\partial_n {\cal F}_{m\mu}=0.\label{bifluc}
\ee
In order to cancel the problematic delta term proportional to $|Q|^2$ in ${\hat F}_{mn}$,
we take the solution for the fluctuation as
\be
{\cal F}_{mn}=rg|Q|^2\Big(\frac{\delta^2(y)}{e_2}-\frac{1}{V}\Big)\epsilon_{mn}.\label{solffmn}
\ee 
Here the bulk constant term comes from the requirement that integrating the left-hand side over
the extra dimensions for the globally well-defined fluctuation ${\cal A}_m$ vanishes. 
We can also see that for the given solution for $A_m$ the Bianchi identity (\ref{bifluc}) can be solved for the globally well-defined ${\cal F}_{m\mu}$. 

Then, inserting the solution (\ref{solffmn}) into eq.~(\ref{decompfmn}) 
with $\langle{\hat F}_{mn}\rangle=q\epsilon_{mn}$,
we obtain the modified field strength as
\be
{\hat F}_{mn}=\Big(q-\frac{rg|Q|^2}{V}\Big)\epsilon_{mn}.
\ee
Therefore, after integrating over the extra dimensions, 
the bulk gauge kinetic term for the modified field strength becomes
\bea
{\cal L}_F&=&-\frac{1}{4}\int d^2y\, e_6 \,e^{\frac{1}{2}\phi}{\hat F}_{mn}{\hat F}^{mn} \nonumber \\
&=&-e_4 \,\frac{V}{2}\,e^{-3\psi+\frac{1}{2}f}\Big(q-\frac{rg|Q|^2}{V}\Big)^2.\label{efff}
\eea
This term is part of the $U(1)_R$ D-term in the 4D effective supergravity that is used
in the text in section 5.1.

\end{document}